\documentclass[a4]{article}
\usepackage{amssymb}
\usepackage{amsmath}

\newtheorem{teiri}{Theorem}
\newtheorem{prop}{Proposition}
\newtheorem{kei}{Corollary}
\newtheorem{rei}{Example}

\newtheorem{algo}{Algorithm}

\title{A Generalization of the Chow-Liu Algorithm and its Application to Statistical Learning}
\author{Joe Suzuki}

\begin{document}
\maketitle 

% If your paper is accepted and the title of your paper is very long, the style will print
% as headings an error message. Use the following command to supply a shorter title of your
% paper so that it can be used as headings.

% If your paper is accepted and the number of authors is large, the style will print
% as headings an error message. Use the following command to supply a shorter version of the authors
% names so that they can be used as headings (for example, use only the surnames)

%\twocolumn[\aistatstitle{A Generalization of the Chow-Liu Algorithm and its Application to Statistical Learning}\aistatsauthor{ Anonymous Author }\aistatsaddress{ Unknown Institution } ]

\section*{Abstract}
We extend the Chow-Liu algorithm for general random variables while
the previous versions only considered finite cases. In particular, this paper
applies the generalization to Suzuki's learning algorithm that generates from data 
forests rather than trees based on the minimum description length by balancing 
the fitness of the data to the forest and the simplicity of the forest.
As a result, we successfully obtain an algorithm when both of the Gaussian and finite
random variables are present.
\section{Introduction}

Learning statistical knowledge from data takes large computation.
For example, constructing a Bayesian network structure expressed by a directed acyclic graph
from data requires exponential time as the number of nodes (attribute values) increases.
We eventually compromise between 
the accuracy and the time complexity of the learning algorithms 
by choosing its approximation to the best solution.
Even in such situations, how to avoid overestimation should be considered.
In this paper, we address how to efficiently estimate the dependency relation among attributes values
by constructing an undirected graph (a Markov network) via the Chow-Liu algorithm \cite{CL}.

The original Chow-Liu algorithm approximates a probability distribution
by a Dendroid distribution expressed by a tree to obtain the best solution
in the sense that the Kullback-Leibler information is the smallest from the original
distribution. The algorithm utilizes the Kruscal algorithm \cite{AHO}: starting with
a finite set $V$ and weights $\{w_{i,j}\}_{i,j\in V, i\not=j}$
\begin{enumerate}
\item $E:=\{\}$
\item ${\cal E}:=\{\{i,j\}|i,j\in V, i\not=j\}$
\item ${\cal E}:={\cal E}\backslash \{\{i,j\}\}$ for $\{i,j\}\in {\cal E}$ maximizing $w_{i,j}$
\item if $(V,E\cup \{\{i,j\}\})$ does not contain a loop, then $E:=E\cup \{\{i,j\}\}$.
\item if ${\cal E}\not=\{\}$, then go to 3., else terminate.
\end{enumerate}
As a result, a tree $(V,E)$ with the maximum value of $\sum_{\{i,j\} \in E}w_{i,j}$ is obtained.
Mutual information $I(i,j)$ of two random variables $X^{(i)}, X^{(j)}$ is used as $w_{i,j}$
in the Chow-Liu algorithm.

For instance, suppose the values of mutual information $I(i,j)$ of pairs of  $X^{(i)},X^{(j)}$ ($i\not=j$)
are given in Table 1. Then, we follow:
\begin{enumerate}
\item Connect ${X^{(1)},X^{(2)}}$ first because $I(1,2)$ is the largest;
\item connect ${X^{(1)},X^{(3)}}$ because $I(1,3)$ is the largest among the unselected; 
\item do not connect ${X^{(2)},X^{(3)}}$ because $I(2,3)$ is the largest among the unselected
but connecting ${X^{(2)},X^{(3)}}$ will make a loop;
\item connect ${X^{(1)},X^{(4)}}$ because $I(1,4)$ is the largest among the unselected;
\item terminate the process because adding any of the remaining candidates will make a loop.
\end{enumerate}

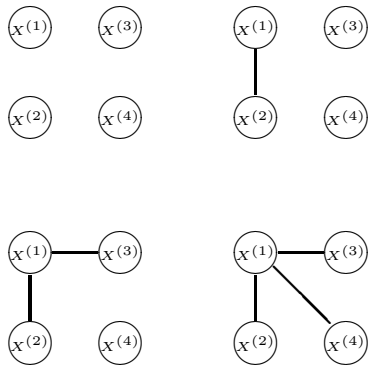
\begin{table}[h]
  \centering
  \caption {Mutual Information for $(i,j)$}
  
  \begin{tabular}{c|c|c}
    \hline
      $i$ & $j$ & $I(i,j)$\\
\hline
       1  &  2  & 12 \\
\hline
       1  &  3  & 10 \\
\hline  
       2  &  3  & 8 \\
\hline
       1  &  4  & 6 \\
\hline
       2  &  4  & 4 \\
\hline
       3  &  4  & 2 \\
\hline
  \end{tabular}
\end{table}
\begin{center}
\tiny
\setlength{\unitlength}{0.30mm}
\begin{picture}(160,160)(-100,-100)
\put(-90,10){\circle{20}}
\put(-50,10){\circle{20}}
\put(-90,50){\circle{20}}
\put(-50,50){\circle{20}}

\put(-90,10){\makebox(0,0){$X^{(2)}$}}
\put(-50,10){\makebox(0,0){$X^{(4)}$}}
\put(-90,50){\makebox(0,0){$X^{(1)}$}}
\put(-50,50){\makebox(0,0){$X^{(3)}$}}

\put(10,10){\circle{20}}
\put(50,10){\circle{20}}
\put(10,50){\circle{20}}
\put(50,50){\circle{20}}

\put(10,10){\makebox(0,0){$X^{(2)}$}}
\put(50,10){\makebox(0,0){$X^{(4)}$}}
\put(10,50){\makebox(0,0){$X^{(1)}$}}
\put(50,50){\makebox(0,0){$X^{(3)}$}}

\put(-90,-90){\circle{20}}
\put(-50,-90){\circle{20}}
\put(-90,-50){\circle{20}}
\put(-50,-50){\circle{20}}

\put(-90,-90){\makebox(0,0){$X^{(2)}$}}
\put(-50,-90){\makebox(0,0){$X^{(4)}$}}
\put(-90,-50){\makebox(0,0){$X^{(1)}$}}
\put(-50,-50){\makebox(0,0){$X^{(3)}$}}

\put(10,-90){\circle{20}}
\put(50,-90){\circle{20}}
\put(10,-50){\circle{20}}
\put(50,-50){\circle{20}}

\put(10,-90){\makebox(0,0){$X^{(2)}$}}
\put(50,-90){\makebox(0,0){$X^{(4)}$}}
\put(10,-50){\makebox(0,0){$X^{(1)}$}}
\put(50,-50){\makebox(0,0){$X^{(3)}$}}

\thicklines
\put(10,20){\line(0,1){20}}

\put(-90,-80){\line(0,1){20}}
\put(-80,-50){\line(1,0){20}}

\put(10,-80){\line(0,1){20}}
\put(20,-50){\line(1,0){20}}
\put(18,-56){\line(1,-1){25}}

%\thinlines
%\put(200,10){\line(1,0){20}}
%\put(230,20){\line(0,1){20}}
%\put(222,44){\line(-1,-1){30}}
\end{picture}
\end{center}

If the distribution is not given but samples are given,
the task is estimation rather than approximation.
Then, the Chow-Liu algorithm uses the maximum likelihood estimators 
of mutual information
 rather than the true mutual information values.
Then, we would only choose a high fitness tree,
without considering the complexity of the trees and the number of parameters:
a (unconnected) forest rather than a (spanning) tree might have been closer to the true distribution.
The order of selecting pairs of nodes may be different if we take into account
the simplicity of the forests/trees structures.

In 1993, Suzuki\cite{Suzuki} proposed a modified version of the Chow-Liu algorithm
 based on the Minimum Description Length in which
 the mutual information is replaced by the one minus a penalty value defined for
  each pair of random variables in order to consider the simplicity of the forest. 
  The modified algorithm obtains the best forest
  in the sense of MDL.

However, those results assume that those random variables take finite values.
This paper deals with the general case: the Chow-Liu and Suzuki algorithms
for general random variables.

In Section 2, we clearly express the Chow-Liu and Suzuki algorithms 
 for capturing essentials. Section 3 deals with the generalizations.
For the Suzuki algorithm, we consider two cases: 
\begin{enumerate}
\item only Gaussian random variables are present.
\item both Gaussian and finite random variables are present.
\end{enumerate}
In Section 4, we summarize the results in this paper and state future works.

\section{For finite random variables}

\subsection{Definitions}
Let $V$ and $E$ be a finite set and a subset of ${\cal E}:=\{\{u,v\}|u,v\in V,u\neq v\}$, respectively.
The pair $(V,E)$ is said an undirected graph. For undirected graph $G=(V,E)$,
$V$ and its elements are said a vertex set and a vertex of $G$, respectively; and
$E$ and its elements are said an edge set and an edge  of $G$, respectively.
The sequence $\{v_i\}_{i=0}^k$ ($k=0,1,\cdots$) is said a path connecting $v_0,v_k \in V$
if there exist $v_1,\cdots,v_{k-1}\in V$ such that
$\{v_{i-1},v_{i}\}\in E,\ i=1,\cdots,k$.
In particular, if $v_0=v_k$, the path $\{U_i\}_{i=0}^k$ is said 
a loop.
The undirected graph $G$ is said a forest if $G$ does not contain any loop,
and is said to be connected if there exists a path connecting each pair of vertexes in $G$.
Any connected forest is said a tree.

On the other hand, a pair  of a finite set $V$ and a subset $\vec{E}$ of 
$\{(u,v)|u,v\in V,u\neq v\}$ is said a directed graph.
In directed graphs, we distinguish $(u,v),(v,u)\in \vec{E}$.

For each $i,j=1,\cdots,N$ ($i\not=j$),
let $X^{(i)}$ be random variables that take finite values in $X^{(i)}(\Omega)$,
$P_i(x)$  a probability of $X^{(i)}=x\in X^{(i)}(\Omega)$, 
$P_{i,j}(x,y)$ a probability of $X^{(i)}=x\in X^{(i)}(\Omega)$ and $X^{(j)}=y\in X^{(j)}(\Omega)$, and
$P_{i\leftarrow j}(x|y)$ a conditional probability $X^{(i)}=x\in X^{(i)}(\Omega)$ given
$X^{(j)}=y\in X^{(j)}(\Omega)$ ($P_i(x), x\in X^{(i)}(\Omega)$ if $j=0$).
We define the mutual information between 
$X^{(i)},X^{(j)}$ by \cite{Cover}
$$I(i,j):=\sum_{x\in X^{(i)}(\Omega), y\in X^{(j)}(\Omega)}
P_{i,j}(x,y)\log \frac{P_{i,j}(x,y)}{P_{i}(x)P_{j}(y)}\ .$$

We assume a natural bijection between $N$ vertexes in $V=\{1,\cdots,N\}$ and $N$ random variables $X^{(1)},\cdots,X^{(N)}$.

\subsection{The original Chow-Liu algorithm}

We consider to approximate the probability $P_{1,\cdots,N}(x^{(1)},\cdots,x^{(N)})$ of
$X^{(1)}=x^{(1)}\in X^{(1)}(\Omega),\cdots,X^{(N)}=x^{(N)}\in X^{(N)}(\Omega)$ by 
\begin{equation}\label{eq27}
Q_{1,\cdots,N}(x^{(1)},\cdots,x^{(N)}):=\prod_{i=1}^N P_{i\leftarrow \pi(i)}(x^{(i)}|x^{(\pi(i)})
\end{equation}
(the Dendroid distribution), where
$\pi: \{1,\cdots,N\}\rightarrow \{0,1,\cdots,N\}$ is to satisfy $\pi^k(i)\not=i$, 
$i=1,\cdots,N$, $k=1,2,\cdots$ if we define
$$\pi^0(i)=i,\ \pi^{k}(i)=\pi(\pi^{k-1}(i)),\ k=1,2,\cdots \ .$$
Although the Dendroid distribution (\ref{eq27}) is expressed by 
a directed graph with emitting vertexes $j\in \{1,\cdots,N\}$ such that $\pi(j)=0$ in general,
it can be regarded as an undirected $(V,E)$ such that
$V:=\{1,\cdots,N\}$ and 
$E:=\{\{i,\pi(i)\}|\pi(i)\not=0, i\in V\}$.
Since 
$$\displaystyle Q_{1,\cdots,N}(x^{(1)},\cdots,x^{(N)})=0 \Longrightarrow P_{1,\cdots,N}(x^{(1)},\cdots,x^{(N)})=0$$
is true, we can define the Kullback-Leibler information from
$P_{1,\cdots,N}$ to $Q_{1,\cdots,N}$ \cite{Cover}:
\begin{eqnarray*}
&&D(P_{1,\cdots,N}||Q_{1,\cdots,N})\\
&:=&\sum_{x^{(1)}\in X^{(1)}(\Omega), \cdots, x^{(N)}\in X^{(N)}(\Omega)}
P_{1,\cdots,N}(x^{(1)},\cdots,x^{(N)})\\
&&\cdot \log \frac{P_{1,\cdots,N}(x^{(1)},\cdots,x^{(N)})}{Q_{1,\cdots,N}(x^{(1)},\cdots,x^{(N)})}\ .
\end{eqnarray*}
We wish to identify $Q_{1,\cdots,N}$ so that
 the value of $D(P_{1,\cdots,N}||Q_{1,\cdots,N})$ is minimized.
In other words, we evaluate the error by $D(P_{1,\cdots,N}||Q_{1,\cdots,N})$
when we approximate $P_{1,\cdots,N}$ by $Q_{1,\cdots,N}$, and find $\pi$ minimizing it.
On the other hand, since
\begin{eqnarray}
&&Q_{1,\cdots,N}(x^{(1)},\cdots,x^{(N)})\nonumber\\
&=& 
\{\prod_{\pi(j)=0}P_{j}(x^{(j)})\}\cdot
\{\prod_{\pi(i)\not=0} \frac{P_{i,\pi(i)}(x^{(i)},x^{(\pi(i))})}{P_{\pi(i)}(x^{(\pi(i))})}\}\nonumber \\
&=&\{\prod_{\pi(i)\not=0} \frac{P_{i,\pi(i)}(x^{(i)},x^{(\pi(i))})}{P_{i}(x^{(i)})P_{\pi(i)}(x^{(\pi(i))})}\}
\cdot \{\prod_{j=1}^N P_{j}(x^{(j)})\}\nonumber\\
&&\label{eq289}\ ,
\end{eqnarray}
we have
\begin{eqnarray}\label{eq267}
&&D(P_{1,\cdots,N}||Q_{1,\cdots,N})=-\sum_{\pi(i)\not=0}I(i,\pi(i))\nonumber\\
&&+\sum_{x^{(1)}\in X^{(1)}(\Omega), \cdots, x^{(N)}\in X^{(N)}(\Omega)} P_{1,\cdots,N}(x^{(1)},\cdots,x^{(N)})\nonumber\\
&&\cdot \log \frac{P_{1,\cdots,N}(x^{(1)},\cdots,x^{(N)})}{\prod_{i=1}^NP_{i}(x^{(i)})}\ .
\end{eqnarray}
to find the last term in (\ref{eq267}) does not depend on $\pi$.
Hence, minimizing $D(P_{1,\cdots,N}||\tilde{P}_{1,\cdots,N})$ is equivalent to
maximizing $\sum_{\{i,j\}\in E}I(i,j)$.
In this case, the (undirected) forest has only one $i\in V$ such that $\pi(i)=0$ (undirected tree).

To this end, we apply the Kruscal algorithm which is used for maximizing the total weights
along with the obtained tree if we have the values of weights for all the pairs of vertexes beforehand.
In this case, the value of each edge is the mutual information $I(i,j)$:
\begin{algo}[Chow-Liu, 1968] \rm　
\begin{description}
\item[Input] $\{I(i,j)\}_{i\not=j}$
\item[Output] $E$
\end{description}
\begin{enumerate}
\item $E:=\{\}$;
\item ${\cal E}:=\{\{i,j\}|i\not=j\}$;
\item ${\cal E}:={\cal E}\backslash \{\{i,j\}\}$ for $\{i,j\}\in {\cal E}$ maximizing $I(i,j)$;
\item if $(V,E\cup \{\{i,j\}\})$ does not contain loop, then $E:=E\cup \{\{i,j\}\}$;
\item if ${\cal E}\not=\{\}$, then go to 3, else terminate.
\end{enumerate}
\end{algo}
($\cup$ and $\backslash$ denote the addition and subtraction of two sets.)

The Kruscal algorithm outputs a tree with the maximum total weights (Aho, Hopcraft, Ullman, 1974 \cite{AHO}).

\subsection{Maximizing Likelihood}

If distributions such as 
$P_{1,\cdots,N}$, $Q_{1,\cdots,N}$ are not given, we need to estimate the parameters $\theta$ expressing 
%\begin{equation}\label{eq278}
$P(x^{(1)},\cdots,x^{(N)}|\theta)$ and $Q(x^{(1)},\cdots,x^{(N)}|\theta)$.
%\end{equation}
In this case,  if we differentiate 
$-\log P(x^{(1)},\cdots,x^{(N)}|\theta)$
by each component of $\theta$ to obtain the maximum likelihood estimators $\hat{\theta}$,
we find that they are relative frequencies:
$$P(x^{(1)},\cdots,x^{(N)}|\hat{\theta}(x^n))=\frac{c_{1,\cdots,N}(x^{(1)},\cdots,x^{(N)})}{n}\ ,$$
where $c_{1,\cdots,N}(x^{(1)},\cdots,x^{(N)})$ is the numbers of occurrences of
$(X^{(1)},\cdots,X^{(N)})=(x^{(1)},\cdots,x^{(N)})\in X^{(1)}(\Omega)\times \cdots \times  X^{(N)}(\Omega)$.

Given $n$ training sequences
$$x^n:=\{(x_i^{(1)},\cdots,x_i^{(N)})\}_{i=1}^n\in (X^{(1)}(\Omega)\times \cdots \times X^{(N)}(\Omega))^n\ ,$$
let 
$c_{i}(x)$,
$c_{j}(y)$, and
$c_{i,j}(x,y)$
be the numbers of occurrences of 
$X^{(i)}=x\in X^{(i)}(\Omega)$,
$X^{(j)}=y\in X^{(i)}(\Omega)$, and
$(X^{(i)},X^{(j)})=(x,y)\in X^{(i)}(\Omega)\times X^{(j)}(\Omega)$, respectively.
Then, minimizing 
\begin{eqnarray*}
&&D(P(\cdot|\hat{\theta}(x^n))||Q(\cdot|\hat{\theta}(x^n)))\\
&=&\sum_{x^{(1)}\in X^{(1)}(\Omega),\cdots,x^{(N)}\in X^{(N)}(\Omega)}\\
&&P(x^{(1)},\cdots,x^{(N)}|\hat{\theta}(x^n))
\log \frac{P(x^{(1)},\cdots,x^{(N)}|\hat{\theta}(x^n))}{Q(x^{(1)},\cdots,x^{(N)}|\hat{\theta}(x^n))}
\end{eqnarray*}
is equivalent to minimizing 
\begin{eqnarray}
H(\pi,x^n)&:=&\sum_{i=1}^n -\log Q(x_i^{(1)},\cdots,x_i^{(N)}|\hat{\theta}(x^n))\nonumber\\
&=&
-\sum_{\{i,j\}\in E}I_n(i,j)\nonumber\\
&&+n\sum_{i=1}^N\sum_{x^{(i)}\in X^{(i)}(\Omega)} -P_{i}(x^{(i)}|\hat{\theta}(x^n))\log {P_{i}(x^{(i)}|\hat{\theta}(x^n))}\ .
\label{eq245}
\end{eqnarray}
and $I(i,j)$ in Algorithm 1 is replaced by 
\begin{eqnarray*}
&&I_n(i,j)\\
&:=n&\sum_{x\in X^{(i)}(\Omega), y\in X^{(j)}(\Omega)}
P_{i,j}(x,y|\hat{\theta})\log \frac{P_{i,j}(x,y|\hat{\theta}(x^n))}{P_{i}(x|\hat{\theta}(x^n))P_{j}(y|\hat{\theta}(x^n))}\\
&=& 
\sum_{x\in X^{(i)}(\Omega), y \in X^{(j)}(\Omega)}
{c_{i,j}(x,y)}\log \frac{c_{i,j}(x,y)}{c_{i}(x)c_{j}(y)}
\end{eqnarray*}
to obtain the structure $\pi$ for the Dendroid distribution.

More accurate learning results could be obtained without approximating to 
the Dendroid distribution, say depending on more than one parent.
However, exponential order computation of $N$ is required in general.
The Chow-Liu algorithm and its variant 
complete in  $O(N^2)$ time, and is easier to apply to realistic problems.

\subsection{Minimizing description length}

Another way to deal with the case that distributions $P_{1,\cdots,N}, Q_{1,\cdots,N}$ are not given
is to mixture 
$P(x^{(1)},\cdots,x^{(N)}|\theta)$ and $Q(x^{(1)},\cdots,x^{(N)}|\theta)$
 by $w$ w.r.t. $\theta$ such that $\int w(\theta)d\theta=1$:
\begin{eqnarray*}
P(x^{(1)},\cdots,x^{(N)}):=\int P(x^{(1)},\cdots,x^{(N)}|\theta)w(\theta)d\theta
\end{eqnarray*}
and
\begin{eqnarray*}
Q(x^{(1)},\cdots,x^{(N)}):=\int Q(x^{(1)},\cdots,x^{(N)}|\theta)w(\theta)d\theta\ .
\end{eqnarray*}
We consider to find the structure $\pi$ maximizing 
$$\prod_{i=1}^n Q(x_i^{(1)},\cdots,x_i^{(N)})$$
 or, equivalently, minimizing $\sum_{i=1}^n -\log Q_i(x^{(1)},\cdots,x_i^{(N)})$ rather than minimizing $H(\pi,x^n)$.
The quantity is said description length because it satisfies the Kraft inequality in information theory \cite{Cover}.

Let $\alpha^{(i)}$ be the number of elements in $X^{(i)}(\Omega)$, $i=1,\cdots,N$, and $\alpha^{(0)}:=1$.
We notice that $Q_{1,\cdots,N}$ has $k:=\sum_{i=1}^N(\alpha^{(i)}-1)\alpha^{(\pi(i))}$ parameters:
for each $X^{(\pi(i))}=x^{(\pi(i))}\in X^{(\pi(i))}(\Omega)$,
the probabilities of $X^{(i)}=x^{(i)}\in X^{(i)}(\Omega)$
should be specified.
Then, there exists a constant $C$ such that \cite{Suzuki}
\begin{equation}\label{eq785}
L(\pi,x^n):=H(\pi,x^n)+\frac{k}{2}\log n+C \geq \sum_{i=1}^n -\log Q(x_i^{(1)},\cdots,x_i^{(N)})\ ,
\end{equation}
and the left hand side also satisfies the Kraft inequality for each $\pi$.

The number of parameters increases from $\alpha^{(i)}-1$ to
$\alpha^{(\pi(i))}(\alpha^{(i)}-1)$ if we connect $i$ and $\pi(i)$ as an edge, so that from (\ref{eq245}), 
the description length (\ref{eq785}) becomes 
\begin{eqnarray*}
L(\pi,n)&=&-\sum_{\{i,j\}\in E}I_n(i,j)+\sum_{\{i,j\}\in E}\frac{1}{2}(\alpha^{(i)}-1)(\alpha^{(j)}-1)\log n+C'\ ,
\end{eqnarray*}
where $C'$ is a constant that does not depend on the structure $\pi$.
Thus, we only need to maximize $\sum_{\{i,j\}\in E}J_n(i,j)$ with
\begin{equation}\label{eq703}
J_n(i,j):=I_n(i,j)-\frac{1}{2}(\alpha^{(i)}-1)(\alpha^{(j)}-1)\log n\ .
\end{equation}
This time, we apply the Kruscal algorithm with $\{J_n(i,j)\}_{i\not=j}$ rather than the one with 
$\{I_n(i,j)\}_{i\not=j}$:
\begin{algo}[Suzuki, 1993] \rm　
\begin{description}
\item[Input] $V, \{J_n(i,j)\}_{i\not=j}$
\item[Output] $E$
\end{description}
\begin{enumerate}
\item $E=\{\}$;
\item ${\cal E}:=\{\{i,j\}|i,j\in V,i\not=j\}$;
\item ${\cal E}:={\cal E}\backslash \{\{i,j\}\}$ for $\{i,j\}\in {\cal E}$ maximizing $J_n(i,j)$;
\item If $J_n(i,j)\geq 0$ and $(V,E\cup \{\{i,j\}\})$ does not contain  loop, $E:=E\cup \{\{i,j\}\}$;
\item if ${\cal E}\not=\{\}$,  then go to 3., else terminate
\end{enumerate}
\end{algo}

\begin{rei}\rm
Suppose that the values of 
$J_{n}(i,j)$ are given in Table 2., and that $\alpha^{(1)}=5$, $\alpha^{(2)}=2$, $\alpha^{(3)}=3$, and
$\alpha^{(4)}=4$.
\begin{enumerate}
\item Connect ${X^{(1)},X^{(2)}}$ because $J_{n}(1,2)=8$ is the largest.
\item Connect ${X^{(2)},X^{(3)}}$ because $J_{n}(2,3)=6$ is the largest among the unselected.
\item Do not connect ${X^{(1)},X^{(3)}}$ because $J_{n}(1,3)=2$ is the largest among the unselected but connecting them will make a loop.
\item Connect ${X^{(2)},X^{(4)}}$ because $I_{n}(2,4)=1$ is the largest among the unselected.
\item Terminate the process because for the remaining candidates $(i,j)$,
$J_n(i,j)< 0$ or adding any of them will make a loop.
\end{enumerate}
\end{rei}

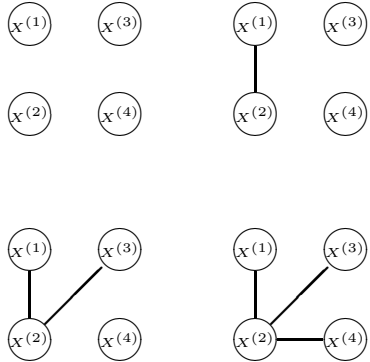
\begin{table}[h]
  \centering
  \caption {Example 2}
  
  \begin{tabular}{c|c|c|c|c|c}
    \hline
      $i$ & $j$ & $I_{n}(i,j)$&$\alpha^{(i)}$&$\alpha^{(j)}$&$J_{n}(i,j)$\\
\hline
       1  &  2  & 12 & 5 & 2 & 8 \\
\hline
       1  &  3  & 10 & 5 & 3 & 2 \\
\hline  
       2  &  3  & 8 & 2 & 3 & 6 \\
\hline
       1  &  4  & 6 & 5 & 4 & -6 \\
\hline
       2  &  4  & 4 & 2 & 4 & 1 \\
\hline
       3  &  4  & 2 & 3 & 4 & -4 \\
\hline
  \end{tabular}
\end{table}　 

%\small
\setlength{\unitlength}{0.30mm}

\begin{center}
\tiny
\begin{picture}(160,180)(-110,-110)
\put(-90,10){\circle{20}}
\put(-50,10){\circle{20}}
\put(-90,50){\circle{20}}
\put(-50,50){\circle{20}}
\put(-90,10){\makebox(0,0){$X^{(2)}$}}
\put(-50,10){\makebox(0,0){$X^{(4)}$}}
\put(-90,50){\makebox(0,0){$X^{(1)}$}}
\put(-50,50){\makebox(0,0){$X^{(3)}$}}

\put(10,10){\circle{20}}
\put(50,10){\circle{20}}
\put(10,50){\circle{20}}
\put(50,50){\circle{20}}
\put(10,10){\makebox(0,0){$X^{(2)}$}}
\put(50,10){\makebox(0,0){$X^{(4)}$}}
\put(10,50){\makebox(0,0){$X^{(1)}$}}
\put(50,50){\makebox(0,0){$X^{(3)}$}}
{\thicklines
\put(10,20){\line(0,1){20}}
}

\put(-90,-90){\circle{20}}
\put(-90,-50){\circle{20}}
\put(-50,-90){\circle{20}}
\put(-50,-50){\circle{20}}
\put(-90,-90){\makebox(0,0){$X^{(2)}$}}
\put(-90,-50){\makebox(0,0){$X^{(1)}$}}
\put(-50,-90){\makebox(0,0){$X^{(4)}$}}
\put(-50,-50){\makebox(0,0){$X^{(3)}$}}
{\thicklines
\put(-90,-80){\line(0,1){20}}
\put(-83,-83){\line(1,1){25}}
}

\put(10,-90){\circle{20}}
\put(50,-50){\circle{20}}
\put(10,-50){\circle{20}}
\put(50,-90){\circle{20}}
\put(10,-90){\makebox(0,0){$X^{(2)}$}}
\put(50,-90){\makebox(0,0){$X^{(4)}$}}
\put(10,-50){\makebox(0,0){$X^{(1)}$}}
\put(50,-50){\makebox(0,0){$X^{(3)}$}}

\thicklines
\put(10,-80){\line(0,1){20}}
\put(17,-83){\line(1,1){25}}
\put(20,-90){\line(1,0){20}}

\end{picture}
\end{center}

Both of $I_n(i,j)$ and $J_n(i,j)$ are criteria for choosing $\{i,j\}$.
We notice that $I_n(i,j)$ only sees if the training sequence $x^n$ fits the structure $\pi$.
On the other hand, $J_n(i,j)$ looks at the simplicity of the forest as well as the fitness,
so that even if ${\cal E}\not=\{\}$, the process stops 
if $J_n(i,j)<0$ for all the rest of $\{i,j\}$'s.
The resulting forest can be either connected or unconnected.
Since the selecting order is different between $\{I_n(i,j)\}_{i\not=j}$ and $\{J_n(i,j)\}_{i\not=j}$,
the structures of the resulting forests are different when the both algorithms complete.

Furthermore, $\displaystyle\frac{k}{2}\log n$ in 
(\ref{eq785}) can be replaced by $\displaystyle\frac{k}{2}d_n$ with nonnegative real sequence 
$\{d_n\}_{n=1}^\infty$
such that $\displaystyle \lim_{n\rightarrow \infty}\frac{d_n}{n}=0$
for general information criteria.

\section{For general random variables}

Consider the general random variables: 
\begin{rei}\rm
Suppose that random variable 
$X$ has the distribution function
\begin{eqnarray*}
F_X(x)=
\left\{
\begin{array}{ll}
0,&x<-1\\
\displaystyle \frac{1}{2},&-1\leq x < 0\\
\displaystyle \frac{1}{2}\int_{0}^x g(t)dt,&0\leq x
\end{array}
\right.\ ,
\end{eqnarray*}
where $\int_0^\infty g(t)=1$.
Such an $X$ does not have any probability density function $f_X$ such that
$F_X(x)=\int_{-\infty}^xf_X(t)dt$, which means $X$ is neither discrete or continuous.
\end{rei}
In this section, how the Chow-Liu and its variants can be extended for 
such general random variables.

\subsection{Definitions}

We fix a probability space $(\Omega, {\cal F}, \mu)$, where
$\Omega$ is a sample space, $\cal F$ is a $\sigma$ set field of $\Omega$, i.e.
a set consisting of the sets obtained by applying 
a countable number of set operations $\cup, \backslash, \cap$
to subsets of $\Omega$.
The elements of $\cal F$ is said an event.
We denote by ${\cal B}$ 
 the $\sigma$ set field generated by the whole open sets in ${\mathbb R}$ (the Borel set field of ${\mathbb R}$).
In general, if the mapping 
$f:\Omega\rightarrow {\mathbb R}$ satisfies
$$D\in {\cal B} \Longrightarrow \{\omega\in \Omega|f(\omega)\in D\}\in {\cal F}\ ,$$
$f$ is said measurable on $\cal F$. The mapping $\nu:{\cal F}\rightarrow {\mathbb R}$
satisfying
\begin{enumerate}
\item $\nu(A)\geq 0, A\in {\cal F}$
\item $A\cap B=\{\} \Longrightarrow \nu(A\cup B)=\nu(A)+\nu(B)$
\item $\nu(\{\})=0$
\end{enumerate}
is said to be a measure. The $\mu$ in the probability space is a measure
such that $\mu(\Omega)=1$ (probability measure).

We can define the Lebesgue integral
$$\int_{A} fd\nu:= \sup_{\{A_i\}}\sum_{i}\{\inf_{\omega\in A_i} f(\omega)\nu(A_i)\}=
\inf_{\{A_i\}}\sum_{i}\{\sup_{\omega\in A_i} f(\omega)\nu(A_i)\}$$
w.r.t. measure $\nu: {\cal F}\rightarrow {\mathbb R}$
and measureble bounded $f$ on ${\cal F}$, where 
$A=\cup_iA_i$, $A_i\cap A_j=\{\}$($i\not=j$).

For measures $\mu,\nu$ on $\cal F$ and $A\in {\cal F}$, if $\nu(A)=0\Longrightarrow \mu(A)=0$,
$\mu$ is said to be absolutely continuous w.r.t. $\nu$, and write $\mu<<\nu$.
Also, we say that measure $\nu$ is $\sigma$-finite if $\Omega=\cup_iA_i$ and $\nu(A_i)<\infty$.
\begin{prop}[Radon-Nikodym]
For each $A\in {\cal F}$, if $\mu, \nu$ are $\sigma$-finite and $\mu<<\nu$,
then there exists measurable $\displaystyle \frac{d\mu}{d\nu}:=f\geq 0$ on ${\cal F}$
such that $$\mu(A)=\int_A fd\nu$$.
\end{prop}
\begin{kei}\rm If $\mu<<\nu<<\lambda$,
$$\frac{d\mu}{d\lambda}=\frac{d\mu}{d\nu}\cdot \frac{d\nu}{d\lambda}$$
\end{kei}

When $\mu<<\nu$,
we define the Kullback-Leibler information
$$D(\mu||\nu):=\int \log(\frac{d\mu}{d\nu})d\mu\ .$$
Properties such as $D(\mu||\nu)\geq 0$, $D(\mu||\nu)=0\Longleftrightarrow \mu=\nu$ are available.

\subsection{Generalization}

In $(\Omega,{\cal F},\mu)$, any measurable mapping $X:\Omega\rightarrow {\mathbb R}$ on 
$\cal F$ is said a random variable.
For 
$D,D'\in {\cal B}$, let
$$\mu_X(D):=\mu(\{\omega\in \Omega|X(\omega)\in D\})\ ,$$
$$\mu_{Y}(D):=\mu(\{\omega\in \Omega|Y(\omega)\in D\})\ ,$$
$$\mu_{XY}(D,D'):=\mu(\{\omega\in \Omega|X(\omega)\in D, Y(\omega)\in D' \})\ ,$$
$$\mu_{X|Y}(D,D'|D):=\frac{\mu_{XY}(D,D')}{\mu_Y(D)}\ {\rm for}\ {\mu_Y(D)}>0\ ,$$
and
$$\nu_{XY}(D,D'):=\mu_X(D)\mu_Y(D')\ .$$
%So, we define the quantity 

Then, we have
$$\nu_{XY}(D,D')=0 \Longrightarrow \mu_{XY}(D,D')=0\ ,$$
which means that $\mu_{XY}$ is absolutely continuous  w.r.t. $\nu_{XY}$.
We define the mutual information between $X,Y$
 by
$$I(X,Y):=D(\mu_{XY}||\nu_{XY})=
\int_{x\in X(\Omega), y\in Y(\Omega)}
\mu_{XY}(dx,dy)
\log \frac{d\mu_{XY}}{d{\nu}_{XY}}(x,y)\ .$$
Hereafter, we denote $\displaystyle \frac{d\mu_{XY}}{d\nu_{XY}}$ in the definition by
$\displaystyle \frac{d^2\mu_{XY}}{d\mu_Xd\mu_Y}$.

%From Radon-Nykodim, 
%there exists measurable $f:{\mathbb R}\rightarrow {\mathbb R}_{\geq 0}$ 
%on $\cal B$ such that
%$$\mu((X\in D) \cap (Y\in D'))=\int_{(Y\in D')} f d\mu$$
%for $D,D'\in {\cal B}$.
%これを事象$(Y\in D')$のもとでの$X$の条件付測度と呼び、$\mu_{X|Y}(D|D')$とかく。
%以下では、一般性を失うことなく、$\Omega={\mathbb R}, {\cal F}={\cal B}$を仮定する。

For random variables $X^{(1)},\cdots,X^{(N)}$, we define
$\mu_i(D):=\mu_{X^{(i)}}(D)$, 
$\mu_{i,j}(D,D'):=\mu_{X^{(i)},X^{(j)}}(D,D')$, $I(i,j):=I(X^{(i)},X^{(j)})$, and
$\mu_{i\leftarrow j}(D|D'):=\mu_{X^{(i)}|X^{(j)}}(D|D')$ for $i, j=1,\cdots,N$ ($i\not=j$), 
and $\mu_{i\leftarrow j}(D|D'):=\mu_i(D)$ if $j=0$.

For $D^{(1)},\cdots,D^{(N)}\in {\cal B}$,
we approximate 
$$\mu_{1,\cdots,N}(D^{(1)},\cdots,D^{(N)}):=\mu(\{\omega\in \Omega|X^{(1)}(\omega)\in D^{(1)},\cdots,X^{(N)}(\omega)\in D^{(N)}\})$$
by
\begin{equation}\label{eq272}
%d\tilde{\mu}_{1,\cdots,N}(x_1,\cdots,x_N)=
{\nu}_{1,\cdots,N}(D^{(1)},\cdots,D^{(N)}):=\prod_{i=1}^N \mu_{i\leftarrow \pi(i)}(D^{(i)}|D^{(\pi(i))})\ .
\end{equation}
From 
$$\displaystyle {\nu}_{1,\cdots,N}(D^{(1)},\cdots,D^{(N)})=0 \Longrightarrow \mu_{1,\cdots,N}(D^{(1)},\cdots,D^{(N)})=0\ ,$$
the Kullback-Leibler information $\mu_{1,\cdots,N}$ w.r.t. ${\nu}_{1,\cdots,N}$ is defined:
\begin{eqnarray*}
&&D(\mu_{1,\cdots,N}||\tilde{\mu}_{1,\cdots,N}):=\int_{x^{(1)}\in X^{(1)}(\Omega), \cdots, x^{(N)}\in X^{(N)}(\Omega)}\\
&&\mu_{1,\cdots,N}(dx^{(1)},\cdots,dx^{(N)})
\log \frac{d\mu_{1,\cdots,N}}{d{\nu}_{1,\cdots,N}}(x^{(1)},\cdots,x^{(N)})
%\log \frac{d\mu_{1,\cdots,N}(x^{(1)},\cdots,x^{(N)})}{d\tilde{\mu}_{1,\cdots,N}(x^{(1)},\cdots,x^{(N)})}
\end{eqnarray*}
We wish to find ${\nu}_{1,\cdots,N}$ such that $D(\mu_{1,\cdots,N}||{\nu}_{1,\cdots,N})$ is minimized.

\begin{teiri}
There exists a constant $C$ not depending on $\pi$ such that 
\begin{eqnarray*}
D(\mu_{1,\cdots,N}||{\nu}_{1,\cdots,N})=-\sum_{\pi(i)\not=0}I(i,\pi(i))+C\ .
\end{eqnarray*}
\end{teiri}
Proof: Generalizing (\ref{eq289}), from (\ref{eq272}), we have
\begin{eqnarray}
&&{\nu}_{1,\cdots,N}(D^{(1)},\cdots,D^{(N)})\nonumber\\
&=& 
\{\prod_{\pi(j)=0}\mu_{j}(D^{(j)})\}\cdot
\{\prod_{\pi(i)\not=0} \frac{\mu_{i,\pi(i)}(D^{(i)},D^{(\pi(i))})}{\mu_{\pi(i)}(D^{(\pi(i))})}\}\nonumber \\
&=&
\{\prod_{\pi(i)\not=0} \frac{\mu_{i,\pi(i)}(D^{(i)},D^{(\pi(i))})}{\mu_{i}(D^{(i)})\mu_{\pi(i)}(D^{(\pi(i))})}\}
\cdot \{\prod_{j=1}^N \mu_{j}(D^{(j)})\}\ .\label{eq889}
\end{eqnarray}
Let
$$\eta_{1,\cdots,N}(D^{(1)},\cdots,D^{(N)}):=\prod_{j=1}^N\mu_j(D^{(j)})\ .$$
Then, we have
\begin{equation}\label{eq8}
\frac{d\nu}{d\eta}=\prod_{\pi(i)\not=0}\frac{d^2\mu_{i,\pi(i)}}{d\mu_id\mu_{\pi(i)}}
\end{equation}
(see Appendix for proof). From the corollary, we have
$$\frac{d\mu}{d\nu}=
\frac{d\mu}{d\eta}/\frac{d\nu}{d\eta}=
[\prod_{\pi(i)\not=0}\frac{d^2\mu_{i,\pi(i)}}{d\mu_id\mu_{\pi(i)}}]^{-1}\frac{d\mu}{d\eta}\ .$$
Furthermore, taking $E\log$ for the both sides, we have
\begin{eqnarray*}
&&E\log \frac{d\mu}{d\nu}(X^{(1)},\cdots,X^{(N)})\\
&=&
-\sum_{\pi(i)\not=0}
E\log \frac{d^2\mu_{i,\pi(i)}}{d\mu_id\mu_{\pi(i)}}(X^{(i)},X^{(\pi(i))})+E\log \frac{d\mu}{d\eta}(X^{(1)},\cdots,X^{(N)})\ .
\end{eqnarray*}
This completes the proof.

\subsection{When only Gaussian random variables are present}

We express the probability density funcions of $X^{(i)}\sim N(\mu^{(i)},\sigma^{2}_{ii})$
and  $(X^{(i)},X^{(j)})\sim {\cal N}((\mu^{(i)},\mu^{(j)}),\Sigma)$ by
\begin{equation}
f_{X^{(i)}}(x^{(i)}):=\frac{1}{\sqrt{2\pi {\sigma_{ii}}}}\exp\{-\frac{(x^{(i)}-\mu^{(i)})^{2}}{{2\sigma_{ii}}} \}\nonumber
\end{equation}
and
\begin{equation}
f_{X^{(i)}X^{(j)}}(x^{(i)},x^{(j)}):=\frac{1}{2\pi {|\Sigma|}^{\frac{1}{2}}}
\exp\{-\frac{1}{2}{{}^t({x^{(i)}-\mu^{(i)}}, {x^{(j)}-\mu^{(j)}})}{\Sigma^{-1}}{({x^{(i)}-\mu^{(i)}}, {x^{(j)}-\mu^{(j)}})}\}\nonumber\ ,
\end{equation}
respectevly, where
$\Sigma=\begin{pmatrix} {{\sigma}_{ii}}  & {{\sigma}_{ij}} \\ {{\sigma}_{ji}} & {{\sigma}_{jj}} \end{pmatrix}$.
Let $\displaystyle \rho_{i,j}:=\frac{\sigma_{ij}}{\sqrt{{\sigma_{ii}}{\sigma_{jj}}}}$ be the correlation factor.
Then, $I(i,j)$ can be obtained via $\rho_{ij}$:
\begin{eqnarray}
I(i,j)&=&\int \int f_{X^{(i)}X^{(j)}}(x^{(i)},x^{(j)})
\log {\frac{f_{X^{(i)}X^{(j)}}(x^{(i)},x^{(j)})}{f_{X^{(i)}}(x^{(i)})f_{X^{(j)}}(x^{(j)})}}dx^{(i)}dx^{(j)}\nonumber \\
&=&\log {\frac{\sqrt{{{\sigma_{ii}}}{{\sigma_{jj}}}}}{{|\Sigma|}^{\frac{1}{2}}}}\nonumber \\
&=&-\frac{1}{2}\log{(1-{\rho_{ij}}^{2})}\nonumber\ .
\end{eqnarray}
Chow-Liu algorithm can be applied using those values.

As obtained in Section 2.3, the maximum likelihood estimators of $I_n(i,j)$
$$\hat{\mu}^{(i)}=\frac{1}{n}\sum_{h=1}^nx_h^{(i)}$$
$$\hat{\rho}_{i,j}=\frac{\sum_{h=1}^n(x_h^{(i)}-\hat{\mu}^{(i)})(x_h^{(j)}-\hat{\mu}^{(j)})}
{\sqrt{\sum_{h=1}^n(x_h^{(i)}-\hat{\mu}^{(i)})^2\sum_{h=1}^n(x_h^{(j)}-\hat{\mu}^{(j)})^2}}$$
$$I_n(i,j)=-\frac{1}{2}\log{(1-{\hat{\rho}_{ij}}^{2})}$$
 can be 
obtained from the training sequence of length $n$: 
$x^n=\{(x_i^{(1)},\cdots,x_i^{(N)})\}_{i=1}^n\in (X^{(1)}(\Omega)\times \cdots \times X^{(N)}(\Omega))^n$

Let 
$\lambda_{i,j}\in {\mathbb R}$, $X^{(i)}=\epsilon_{i}\sim {\cal N}(0,\phi_i)$,
$X^{(j)}=\lambda_{i,j} X^{(i)}+\epsilon_j$, $\epsilon_j\sim {\cal N}(0,\phi_j)$.
Then, we have
$$\Sigma=\left(
\begin{array}{cc}
\phi_{i}&\lambda_{i,j}\phi_{i}\\
\lambda_{i,j}\phi_{i}&\lambda_{i,j}^2\phi_{i}+\phi_{j}\\
\end{array}
\right)\ ,$$
$$\rho_{i,j}=\frac{1}{\sqrt{1+\phi_j/(\lambda_{i,j}^2\phi_i)}}\ ,$$
and
$$\phi_j\lambda_{ji}=\phi_i\lambda_{ij}\ .$$
Thus, $\rho_{i,j},\lambda_{i,j},\lambda_{j,i}$ are bijection among any of two.
Although under the condition
$$\rho_{i,j}=0 \Longleftrightarrow \lambda_{i,j}=0\Longleftrightarrow \sigma_{ii}=\phi_i, \sigma_{jj}=\phi_j, \sigma_{i,j}=0\ ,$$
there are two independent parameters $\sigma_{ii}=\phi_i,\sigma_{jj}=\phi_j$,
if $\lambda_{i,j}\not=0$, another parameter 
$\sigma_{i,j}=\lambda_{i,j}^2\phi_{i,j}+\phi_{i}$ should be specified.

Thus, if we consider the complexity of forests,
adding one edge leads to adding one parameter, so that
$$J_n(i,j)=I_{n}(i,j)-\frac{1}{2}d_n\ .$$
It is possible that the process terminates before the forest becomes a tree
if all the values of the rest of $J_n(i,j)$ are negative then.
However, the order of selecting the edges are the same
for $\{I_n(i,j)\}_{i\not=j}$ and $\{J_n(i,j)\}_{i\not=j}$.

\subsection{When both Gaussian  and finite random variables are present}
We consider the case that both Gaussian and finite random variables are present.
Suppose that $X^{(i)}$ and $X^{(j)}$ are Gussian and finite, respectively.
Then, the mutual information is 
$$I(i,j)=\sum_{y\in X^{(j)}}P_j(y)\int_{x\in X^{(i)}}f_{i\leftarrow j}(x|y)\log \frac{f_{i\leftarrow j}(x|y)}{\sum_{z\in X^{(j)}}P_j(z)f_{i\leftarrow j}(x|z)}dx$$
where $P_j(y):=\mu_Y(\{y\}), y\in X^{(j)}(\Omega)$, and
$f_{i\leftarrow j}(x|y)$
is the conditional Gauss distribution given 
$X^{(j)}=y$.
Thus, $X^{(i)}$ has as many Gaussian distributions as the values $X^{(j)}$ takes.
In particular, if for unknown 
$g: X^{(j)}(\Omega)\rightarrow {\mathbb R}$ and
$\epsilon_i\sim {\cal N}(0,\phi_i)$
$$X^{(i)}=g(X^{(j)})+\epsilon_i$$
\begin{equation}\label{eq251}
f_{i\leftarrow j}(x|y)=\frac{1}{\sqrt{2\pi\phi_i}}\exp \{-\frac{(x-g(y))^2}{2\phi_i}\}\ ,
\end{equation}
then the $|X^{(j)}(\Omega)|=\alpha^{(j)}$ papameters 
 $g(y)$, $y\in X^{(j)}$ should be estimated.
The estimated mutual information becomes
$$I_n(i,j)=\sum_{y\in X^{(j)}}\frac{c_j(y)}{n}
\int_{x\in X^{(i)}}\hat{f}_{i\leftarrow j}(x|y)\log \frac{\hat{f}_{i\leftarrow j}(x|y)}{\sum_{z\in X^{(j)}}\frac{c_j(z)}{n}\hat{f}_{i\leftarrow j}(x|z)}dx\ ,$$
where $\hat{f}_{i\leftrightarrow j}(\cdot|y)$ is the 
estimated probability density function in which  $g(y)$ in 
(\ref{eq251}) is replaced by the maximum likelihood estimator $\hat{g}(y)$: solve
$\partial L/\partial P_j(y)=0$, $\partial L/\partial g(y)=0$, $y\in X^{(j)}(\Omega)$ for 
$$L=\log \prod_{i=1}^n\{f(x_h^{(i)}|g(x_h^{(j)}))P_j(x_h^{(j)})\}+\lambda\{1-\sum_{y\in X^{(j)}(\Omega)}P_j(y)\}$$
to obtain
$$\hat{P}_j(y)=\frac{c_j(y)}{n}$$
$$\hat{g}(y)=\frac{1}{c_j(y)}\sum_{h=1}^nx_h^{(i)}I[x^{(j)}_h=y]\ ,$$
where $I[x^{(j)}_h=y] =1$ if $x^{(j)}_h=y$, and 0 otherwise.

However, if $X^{(i)}$ and $X^{(j)}$ are independent,
then $g$ is a constant and $g(y)=\mu^{(j)}$ for all $y\in X^{(j)}$.
Thus, $$\hat{g}(y)=\frac{1}{n}\sum_{h=1}^nx_h^{(i)}\ .$$

If $\{i,j\}$ are not connected as an edge,
the parameters w.r.t. $X^{(i)}$ is only $\mu^{(i)}$ and $\sigma_{ii}=\phi_i$.
However, if they are connected, we need to estimate
 $g(y), y\in X^{(j)}(\Omega)$ and $\phi_i$,
 so that the number of additional parameters is $\alpha^{(j)}-1$:
$$J_n(i,j):=I_n(i,j)-\frac{(\alpha^{(j)}-1)}{2}d_n$$
in which the difference $J_n(i,j)-I_n(i,j)$ depends on $\{i,j\}$, and the selection order may be different.

As a summary:
\begin{enumerate}
\item if both of $X^{(i)},X^{(i)}$ are finite:
%\begin{equation}\label{eq703}
$\displaystyle J_n(i,j)=I_{n}(i,j)-\frac{(\alpha^{(i)}-1)(\alpha^{(j)}-1)}{2}d_n$
%\end{equation}
\item if both of $X^{(i)},X^{(i)}$ are Gaussian: $\displaystyle J_n(i,j)=I_{n}(i,j)-\frac{1}{2}d_n$
\item if $X^{(i)}$ is Gaussian, and $X^{(j)}$ is finite: $\displaystyle J_n(i,j)=I_{n}(i,j)-\frac{(\alpha^{(j)}-1)}{2}d_n$
\end{enumerate}
Therefore, if $X^{(i)}$ is Gaussian, 
we only need to set $\alpha^{(i)}=2$ in (\ref{eq703})

\section{Concluding Remarks}

We extended the Chow-Liu algorithm for the general random variables,
and considered variants to take into account the complexity of the forest
so that overestimation can be avoided for the general setting.

As a future work, we can further consider ways to avoid overestimation 
for various cases as well as the finite and Gaussian cases.

\section*{Appendix: proof of (\ref{eq8})}

We arbitrarily fix $x^N\in {\mathbb R}^N$ and $\epsilon>0$. 
For each rectangle 
\begin{eqnarray*}
&&(D^{(1)},\cdots,D^{(N)})\subseteq D_\epsilon^N\\
&:=&
\{y^N\in {\mathbb R}^N||\frac{d\nu}{d\eta}(x^N)-\frac{d\nu}{d\eta}(y^N)|<\epsilon, 
|\frac{d^2\mu_{i,\pi(i)}}{d\mu_id\mu_{\pi(i)}}(x^N)-\frac{d^2\mu_{i,\pi(i)}}{d\mu_id\mu_{\pi(i)}}(y^N)|<\epsilon,\ {\rm for}\ \pi(i)\not=0 \}\ ,
\end{eqnarray*}
we have from Radon-Nikodym,
$$\nu(D^{(1)},\cdots,D^{(N)})\geq \inf_{y^N\in D_\epsilon^N}\frac{d\nu}{d\eta}(y^N)\eta(D^{(1)},\cdots,D^{(N)})>\eta(D^{(1)},\cdots,D^{(N)})(\frac{d\nu}{d\eta}(y^N)-\epsilon)$$
and
$$\nu(D^{(1)},\cdots,D^{(N)})\leq \sup_{y^N\in D_\epsilon^N}\frac{d\nu}{d\eta}(y^N)\eta(D^{(1)},\cdots,D^{(N)})<\eta(D^{(1)},\cdots,D^{(N)})(\frac{d\nu}{d\eta}(y^N)+\epsilon)\ ,$$
thus, if  $\eta(D^{(1)},\cdots,D^{(N)})>0$,
$$
\frac{\nu(D^{(1)},\cdots,D^{(N)})}{\eta(D^{(1)},\cdots,D^{(N)})}-\epsilon<\frac{d\nu}{d\eta}({x^N})<
\frac{\nu(D^{(1)},\cdots,D^{(N)})}{\eta(D^{(1)},\cdots,D^{(N)})}+\epsilon\ .
$$
Similarly, for $\pi(i)\not=0$, if $\mu_{i}(D^{(i)})\mu_{\pi(i)}(D^{(\pi(i))})>0$,
$$
 \frac{\mu_{i,\pi(i)}(D^{(i)},D^{(\pi(i))})}{\mu_{i}(D^{(i)})\mu_{\pi(i)}(D^{(\pi(i))})}
-\epsilon<\frac{d^2\mu_{i,\pi(i)}}{d\mu_id\mu_{\pi(i)}}(x^N)<
\frac{\mu_{i,\pi(i)}(D^{(i)},D^{(\pi(i))})}{\mu_{i}(D^{(i)})\mu_{\pi(i)}(D^{(\pi(i))})}
+\epsilon\ .
$$
Since $x^N\in {\mathbb R}^N$ and $\epsilon>0$ are arbitrary,  (\ref{eq889}) means (\ref{eq8}).
(We only need to consider $x^N\in {\mathbb R}^N$ such that there exists $(D^{(1)},\cdots,D^{(N)})\ni x^N$
satisfying 
$\eta(D^{(1)},\cdots,D^{(N)})>0$ and $\mu_{i}(D^{(i)})\mu_{\pi(i)}(D^{(\pi(i))})>0$
 for all $\epsilon>0$.)

\end{document}